\documentstyle[prl,aps,twocolumn,amsmath,amssymb]{revtex}
\begin{document}
\def\al{\alpha}
\newcommand{\ket}[1]{\left| #1\right>}
\def\PLB{ Phys. Lett.  }
\def\PLA{ Phys. Lett.  }
\preprint{\vbox{\noindent
\null\hfill INFNCA-TH00018\\ IFUM-670-FT}}
\draft
\title{Large {N} limit of Calogero--Moser models and Conformal Field Theories}
\author{Mariano Cadoni$^{a,}$\thanks{E--Mail: cadoni@ca.infn.it},
Paolo Carta$^{a,}$\thanks{E--Mail: carta@ca.infn.it} and Dietmar
Klemm$^{b,}$\thanks{E--Mail: dietmar.klemm@mi.infn.it}}
\address{$^a$ Universit\`a degli Studi di Cagliari, Dipartimento di
Fisica and\\INFN, Sezione di Cagliari, Cittadella Universitaria 09042,
Monserrato, Italy.}  \address{$^b$ Universit\`a degli Studi di Milano,
Dipartimento di Fisica and\\ INFN, Sezione di Milano, Via Celoria 16,
20133 Milano, Italy.}
\maketitle
\begin{abstract}
We discuss the large $N$ limit of Calogero--Moser models for the
classical infinite families of simple Lie algebras $A_N$, $B_N$, $C_N$
and $D_N$. We show that the limit defines two different Conformal
Field Theories with central charge $c>1$. The value of $c$ and the
dimension of the  primary field are dictated by the underlying
algebraic symmetries of the model.
\end{abstract}
 %serve anche a creare la pagina del titolo a se.
The classical and quantum integrable model of $N$ interacting bodies
in one spatial dimension introduced by Calogero \cite{cal} has been
studied intensively and still attracts much attention. Surprisingly,
the model and its various generalizations \cite{suth} have been found
relevant to describe properties of theories ranging from spin chains,
quantum Hall effect, two dimensional gravity and random
matrices. Recently, in the high energy and gravity context the
interest was renewed since the proposal \cite{gb} that the model (in a
supersymmetric extension) could be relevant to provide a microscopic
description of the extremal Reissner--Nordstr\"om black hole.

The model is described by the Hamiltonian $H=\sum_{i=1}^N p_i^2/2 +
U(q)$, where $p=(p_1,\ldots,p_N)$ and
$q=(q_1,\ldots,q_N)$ are momentum and coordinate operators in the $N$
dimensional space $\frak{H}$ with $p_l=-i\partial/\partial q_l$
 (particle masses and the Planck constant
are set equal to unity). In the most general formulation of the model
the two--body potential $U(q)$ is defined by  means of the
subsystem $R_+$ of positive roots $\{\al\}$ of a root system defining
a Coxeter group\cite{ol}. 

We will be concerned with the
Calogero--Moser (CM) case,
\begin{equation}\label{e:cm}
U(q)= \sum_{\alpha \in R_+} g^2_\al q^{-2}_\al
+\frac{\omega^2}{2}\sum_{\alpha \in R_+}q^2_\al,
\end{equation}
where $q_\al=(q,\al)$ and the couplings obey the stability condition
\begin{equation}\label{e:stab}
g^2_\al \ge -\frac{|\al|^2}{8},
\end{equation}
$|\al|$ being the length of the root $\al$. Since we are interested in
the large $N$ limit we consider only the infinite (classical) root
systems $A_N$, $B_N$, $C_N$ and $D_N$, corresponding, respectively,
to the simple Lie algebras $su(N+1)$, $so(N+1)$, $sp(N)$ and
$so(2N)$. The only, nontrivial, nonreduced case, $BC_N$, is equivalent to
$A_{2N}$ with proper symmetry conditions preserved by the time
evolution \cite{ol}. Every other reducible case corresponds to the union
of noninteracting systems.

We aim to study the $N\to\infty$ limit  of the model (\ref{e:cm})
pointing out a precise, yet very simple, relation with Conformal Field
Theory (CFT) universality classes, or, more exactly, with, say, the
holomorphic sector of them. We consider a CFT in $D=1+1$ defined on a
strip or on a cylinder. We will show that in the large $N$
limit the CM model  is equivalent  to two different CFTs, 
depending on the root system considered.

We stress that the large $N$ limit we consider here is not the
thermodynamic limit, where the particle density is kept fixed. The
latter limit was considered in previous papers dealing with similar models. In
Ref. \cite{kata} for instance, using finite--size scaling relations, it
was found that the low--energy physics of a Calogero--Sutherland (CS)
model (see Eq. (\ref{e:cs}) below) is described by a $c=1$ CFT. In Ref.
\cite{sciuto}, a second quantized ($D=1+1$) Hamiltonian
has been  studied, which corresponds  to the CS mechanical model. It was shown
that that if one
takes into account the leading ($1/N$) and subleading ($1/N^2$) terms
in the thermodynamic limit,  the dynamics of small
fluctuations around the Fermi surface is described by an effective
field theory endowed with an extended $W_{1+\infty}$ conformal symmetry. 
Our approach is rather different: we do not
consider a thermodynamic limit and will not eventually write down an
effective field theory; on the contrary we use the \emph{entire}
spectrum, not only the lowest--lying exited states, to reconstruct a CFT$_2$
via a large $N$ limit. Needless to say that we rely on the exact
integrability of the CM model, while, for instance, in
Ref. \cite{sciuto} no use is made of the exact form of the spectrum.

The starting point is the very simple form of the spectrum of the CM
model. It is well known that the spectrum is generated by $N$
decoupled harmonic oscillators \cite{ol}. Moreover, it is possible to
define similarity or unitary transformations that map the rational CM
model (\ref{e:cm}) into a set of decoupled harmonic oscillators
\cite{eq}. In this sense the equivalence of the CM model  with a
CFT in the large $N$ limit is self-evident. We want to identify the CFT
precisely.

Let us briefly recall the spectral properties of the CM model (see
Ref. \cite{ol} for a review). Whenever the stability condition
(\ref{e:stab}) is satisfied the Hamiltonian  $H$ is self--adjoint and the
energy level $E_k$ ($H\psi_k=E_k\psi_k$) is  characterized by $N$
integers. If the couplings $g_\al$ are written as usual as $g_\al^2
= \tilde\mu_\al(\tilde\mu_\al-1)|\al|^2/2$, with $\tilde\mu_\al =
\mu_\al + \mu_{2\al}$, the ground state energy $E_0$ is given by
\begin{equation}\label{e:zero}
E_0=\omega\left(\frac{N}{2} + \mu\right), \qquad \mu = \sum_{\alpha \in
R_+}\mu_\al.
\end{equation}
The energies of the excited states read
\begin{equation}
E_{n_1,\ldots,n_N} = E_0 + \omega(\nu_1n_1+\cdots + \nu_Nn_N),
\end{equation} 
where the positive integers $\nu_1,\ldots,\nu_N$ are the degrees of
the $N$ polynomial invariants of the Coxeter group, i.e. the orders
of the Casimir operators of the  Lie algebra under consideration. 
Given the order
of the invariants, it is immediate to compute the quantum partition
function $Z(\beta)$:
\begin{eqnarray}
\lefteqn{Z(\beta) =} \nonumber \\
& & e^{\beta E_0}[(1-e^{-\beta\omega\nu_1})(1-e^{-\beta\omega\nu_2})\cdots
(1-e^{-\beta\omega\nu_N})]^{-1}. \label{e:zcal}
\end{eqnarray}
The orders of the invariants for the classical root systems are known
from the classic work of Racah \cite{racah}, and are quoted in table
I. Neglecting as usual the zero point energy, which is divergent in
the large $N$ limit, see Eq. (\ref{e:zero}), the
partition functions have a very simple form in all cases.

Let us now consider the CFT side. The Hilbert space of a generic
two-dimensional conformal field
theory is given by (see for instance Ref. \cite{dif})
\begin{equation}
\sum_{h,\bar{h}}V(c,h)\otimes \bar{V}(c,\bar{h}),
\end{equation}
where the sum is taken over  the conformal dimensions of the theory and
$c$ is the central charge. $V(c,h)$ is the Verma module generated from
a vacuum $\ket{h}$ with conformal weight $h$, $L_0 \ket{h} = h
\ket{h}$, by the action of the holomorphic Virasoro generators 
$L_n$. $\bar{V}(\bar{h},c)$
describes the antiholomorphic sector (the holomorphic and
antiholomorphic weights are in general independent quantities). The
Verma modules need not to be irreducible representations of the
conformal group. Indeed in general they are reducible. That means
that the states
\begin{equation}\label{e:stati}
L_{-k_1}L_{-k_2}\cdots L_{-k_n}\ket{h}, \qquad 1\le k_1\le \ldots \le
k_n,
\end{equation}
are not necessarily all independent. It may happen that a combination
of the states (\ref{e:stati}) defines a null vector $\ket{\chi}$, that
is a vector annihilated by all of the generators $L_n$ with $n>0$ (we
are considering only the holomorphic sector. For the antiholomorphic
one the same statements hold). A null vector is orthogonal to the whole Verma
module and in particular to himself, $\left<\chi|\chi\right>=0$. This
is of course also true  for all of its descendants. The null submodule
arising from a null state does not contribute to the partition
function, which is computed tracing over an irreducible representation
of the Virasoro algebra. Given the module $V(c,h)$, an irreducible
representation $M(c,h)$ is obtained by quotienting out of it all the null
modules, i.e. identifying two states differing by a null vector.

The null vectors are in correspondence with the zeros of the Kac 
determinant of
the Gram matrix $G$, defined as $G_{ij} = \left< i|j\right>$, where
$\ket{i}$ and $\ket{j}$ are the states (\ref{e:stati}) of the Verma
module $V(c,h)$. The Gram matrix is block diagonal, with blocks
$G^{(l)}$ corresponding to the level $l$, i.e. the $L_0-h$ eigenvalue
of the states (\ref{e:stati}). The expression for the Kac determinant
is (see e.g. Ref. \cite{dif})
\begin{equation}\label{e:kac}
\det G^{(l)} = \al_l\prod_{\substack{r,s\ge l \\ rs \le
l}}[h-h_{r,s}]^{p(l-rs)},
\end{equation}
where $p(l-rs)$ is the number of partitions of the integer $l-rs$ (not
to be confused with the product $p \times (l-rs)$), and $\al_l$ denotes
a positive constant depending on $l$. $h_{r,s}$ are functions of the
central charge $c$ and can be written as
\begin{eqnarray}\label{e:centro}
&&c=13-6\left(t+\frac{1}{t}\right), \nonumber \\ && h_{r,s} =
\frac{1}{4}(r^2-1)t+\frac{1}{4}(s^2-1)\frac{1}{t}-\frac{1}{2}(rs-1),
\end{eqnarray}
where $t$ is a function of $c$. Once we have identified the null vectors
and constructed the irreducible representation $M(c,h)$, the partition
function for the holomorphic sector with conformal weight $h$ can be
defined as
\begin{equation}\label{e:zcft}
Z_h(\beta) = \text{Tr}_{M(c,h)}q^{(L_0-h)},
\end{equation}
where $q\equiv e^{-\beta}$. Notice that the trace is
taken in $M(c,h)$. We bear in mind that $L_0$ is, in proper units, the
energy operator $H$ for $M(c,h)$. In our definition we have again
neglected the zero point energy (basically the vacuum conformal
weight). We note that $Z_h(\beta)$ depends on $h$ since $M(c,h)$ does.
Now the question is if one can define a CFT such that $Z_h(\beta)$
defined by Eq. (\ref{e:zcft}) equals $Z(\beta)$ of Eq. (\ref{e:zcal})
in the limit $N \to \infty$. The answer is simple and affirmative for
the four cases $A_N$, $B_N$, $C_N$ and $D_N$.

$A_N$ is the case of major interest, both from the physical and
mathematical point of view.  We set $\omega =1$. From the table I it follows 
that all the modes but the first are present. 
In order to have the same for the CFT partition function, we simply need a null
vector at the first level. This implies $h=h_{1,1}=0$, because at level
one the only state is $L_{-1}\ket{h}$. The corresponding field
$\phi_{(1,1)}(z)$ is a constant since it satisfy the differential
equation
$$
\frac{\partial}{\partial z}\left< \phi_{(1,1)}(z) X \right > =0,
$$
for every string of fields $X$.  $\phi_{(1,1)}$ is called the identity
operator and denoted by $\Bbb I$. If we choose $c$ such that no more
null vectors are present, then $Z_h(\beta)$ of Eq. (\ref{e:zcft})
equals $Z(\beta)$ given by Eq. (\ref{e:zcal}) since the vectors
$$
L_{-1}\left(L_{-k_1}L_{-k_2}\cdots L_{-k_n}\right)\ket{0}, \qquad 1\le
k_1\le \ldots \le k_n,
$$
do not contribute to the partition function and have the proper
degeneracy (see also Eq. (\ref{e:kac})). It is also evident that the
Hilbert spaces of the two theories have the same structure. The
condition that $h_{r,s}\ne 0$ for $r\ne 1$ and $s\ne 1$ restricts the
possible values of $c$. A general elementary unitary bound on $c$ is
$c\ge 0$ (we consider only unitary theories). From
Eq. (\ref{e:centro}) we immediately see that $c$ cannot be equal to 1,
since in this case $h_{r,r}=0$ $\forall r>0$. On the other hand if
$c<1$ the unitary theories are given by the well known minimal models
whose Hilbert space have a more complicated structure. Indeed the
unitary minimal models with $c<1$ can be labeled by $(p=m+1,p'=m+1)$
with $m=2,3,\ldots$ and satisfy the periodicity property
$h_{r,s}=h_{r+p,s+p'}$ which is sufficient to exclude them. We
conclude that $c>1$ is the only possibility. From Eq. (\ref{e:centro})
it follows  that $t$ as a function of $c$ has two branches and it is given
by
\begin{equation}
t=1+\frac{1}{12}\left[ 1-c\pm\sqrt{(1-c)(25-c)}\right].
\end{equation}
From the previous equation we see that $h_{r,s}$ never vanishes if
$1<c<25$. If $c\ge 25$ we can resort to another well-known expression
of the roots of the Kac determinant, namely
\begin{eqnarray}
&& c= 1-\frac{6}{m(m+1)}, \nonumber \\
&& h_{r,s} = \frac{[(m+1)r-ms]^2-1}{4m(m+1)}.
\end{eqnarray}
If $c\ge 25$ then $-1<m<0$, but this implies that $h_{r,s}$ vanishes only
for $r=s=1$. We conclude that the limit $N\to\infty$ of the $A_N$
systems defines a two-dimensional CFT with $c>1$, $h=0$ and the 
conformal family given by the descendants of $\Bbb I$.

The cases $B_N$, $C_N$ and $D_N$ are less interesting. Setting
$\omega=1/2$ we see from table I that the all the modes $1,2,\ldots$
contribute to $Z(\beta)$ giving the same large $N$ limit for the three
cases. The partition function basically is the character of the
$\widehat{u}(1)$ algebra, the affine extension of $u(1)$, known as the
Heisenberg algebra.  This means that in the limiting CFT the Kac
determinant never vanishes. This condition is ensured taking $h>0$ and
$c>1$. It follows  that we have one conformal
family descendent from the primary field $\ket{h}=\phi(0)\ket{0}$.

It is interesting to note that while the CM model can be defined even for the
noncrystallographic dihedral group $I_2(N)$, with $N=5$, $N\ge 7$, the
large $N$ limit is trivial in this case, since there are only two Casimirs of order
$2$ and $N$.

For high temperatures $\beta\omega \ll 1$, we obtain from the partition
function (\ref{e:zcal}) the energy--temperature relation
in the large $N$ limit,
\begin{equation}
E = E_0 + \frac{\pi^2 T^2}{6\omega}.
\end{equation}
This can be compared with the generic CFT behavior
$E-E_0=\frac{c}{12}\pi LT^2$,
which is valid if the characteristic size $L$ of the system is much larger
than the thermal wavelength of the particles. In this way, we get the equation
\begin{equation}
\frac{2\pi}{\omega} = cL,
\end{equation}
which relates the frequency $\omega$ of the CM model and the product
$cL$ of the central charge and the characteristic size of our two-dimensional 
CFT on
the strip or cylinder.

As a final consideration we note that the procedure used in this letter 
 cannot be employed
for the CS model,
\begin{equation}\label{e:cs}
H = -\frac{1}{2}\sum_{j=1}^N \frac{\partial^2}{\partial x_j^2}
+\gamma(\gamma -1)\frac{\pi^2}{L^2}\sum_{i<j}\frac{1}{\sin^2(\pi(x_i-x_j)/L)},
\end{equation}
where $\gamma$ is a positive constant, often taken to be an
integer. Indeed in this case the difference between two successive
energy levels is proportional to $N$, so that a large $N$ limit is meaningless.

Summarizing, in this letter  we have considered a large $N$ limit 
for the CM model in order
to reconstruct a CFT$_2$ from the exactly known  spectrum of  exited states
of the model.
We have found two different CFTs both with central charge
$c>1$. The value of the central charge as well as the dimension of the
 primary field are dictated only by the algebraic underlying Lie
symmetries of the CM model and are independent of the couplings
strength (provided the stability condition (\ref{e:stab}) is
satisfied).

\vskip 30pt
\begin{table}
\caption{Order of invariants for the for classical root systems $A_n$,
$B_n$, $C_n$ and $D_n$}
\begin{tabular}{ll|l}
\multicolumn{2}{c}{Type of Group} & Order of
invariants\\ \hline
1. $A_n$ & $n\ge 1$ &$2,3,\ldots,(n+1)$ \\
2. $B_n$ & $n\ge 2$ &$2,4,\ldots,2n$ \\
3. $C_n$ & $n\ge 2$ &$2,4,\ldots,2n$ \\
4. $D_n$ & $n\ge 4$ &$2,4,\ldots,(2n-1),n$ \\
\end{tabular}
\end{table}


\begin{references}
\bibitem{cal} F. Calogero, J. Math. Phys. {\bf 10}, 2191 (1969); {\bf
10}, 2191 (1969); {\bf 12}, 419 (1971).
\bibitem{suth} B. Sutherland, J. Math. Phys. {\bf 12}, 246 (1971); {\bf
12}, 251 (1971); Phys. Rev.  {\bf A4}, 2019 (1971); {\bf A5}, 1372
(1972).
\bibitem{gb} G.W. Gibbons and P.K. Townsend, \PLB {\bf B454} (1999) 187.
\bibitem{ol} M.A. Olshanetsky and A.M. Perelomov, Phys. Rep. {\bf 71},
313 (1981); {\bf 94}, 313 (1983).
\bibitem{kata} N. Kawakami and S.-K Yang, Phys. Rev. Lett. {\bf 67},
2493 (1991).
\bibitem{sciuto} M. Frau, A. Lerda, S. Sciuto and G. R. Zemba,
Int. J. Mod. Phys. {\bf A12} 4611 (1997).
\bibitem{eq} N. Gurappa and P.K. Panigrahi, quant--ph/9710019;
cond-mat/9710035.
T. Brzezi\'nski, C. Gonera and P. Ma\'slanka, \PLA {\bf A254} (1999) 185.
\bibitem{racah} G. Racah, Rend. Lincei {\bf 8} (1950) 108.
\bibitem{dif} P. Di Francesco, P. Mathieu and D. S\'en\'echal
\textit{Conformal field theory}, Springer Verlag, 1996. 
\end{references}
\end{document}